\documentclass[aps,prl,showpacs,twocolumn,nofootinbib,preprintnumbers]{revtex4-2}
\usepackage{bbm}
\usepackage{mathrsfs}
\usepackage{epsfig}
\usepackage{soul,xcolor}
\usepackage{graphicx}
\usepackage{amsfonts}
\usepackage{amsthm}
\usepackage[figuresright]{rotating}
\usepackage{amssymb}
\usepackage{amsmath}
\usepackage{dcolumn}
\usepackage{physics}
\usepackage{float}
\usepackage{bm}
\usepackage{physics}
\usepackage{verbatim}
\usepackage{braket}
\usepackage[normalem]{ulem}
\usepackage[ruled,vlined,linesnumbered]{algorithm2e}
\usepackage{setspace}
\usepackage{lipsum}
\usepackage[colorlinks,linkcolor=blue,anchorcolor=blue,citecolor=blue,urlcolor=blue]{hyperref}
\usepackage{enumitem,kantlipsum}

\newcommand\blfootnote[1]{%
  \begingroup
  \renewcommand\thefootnote{}\footnote{#1}%
  \addtocounter{footnote}{-1}%
  \endgroup
}

\begin{document}

\title{Neural Transformer Backflow for Solving Momentum-Resolved Ground States of Strongly Correlated Materials}
\author{Lixing Zhang$^{1}$, Di Luo$^{2*}$\thanks{email: \url{diluo1000@gmail.com}}}

\affiliation{$^{1}$\mbox{Department of Chemistry and Biochemistry, University of California Los Angeles, Los Angeles, CA 90095, USA}\\
$^{2}$\mbox{Department of Electrical and Computer Engineering, University of California Los Angeles, Los Angeles, CA 90095, USA} \\
}
\begin{abstract}
Strongly correlated materials host a rich variety of exotic quantum phases but remain challenging to solve due to strong interactions. We introduce the Neural Transformer Backflow (NTB) framework, a powerful neural-network ansatz formulated within a multi-band projection formalism. NTB is mean-field transcendental, parameter-efficient and fermionic intrinsic, exhibiting superior performance compared with existing neural ansatzes. By naturally enforcing momentum conservation, NTB enables direct computation of momentum-resolved many-body ground states, providing detailed access to degeneracies and energy gaps. It achieves high accuracy on small systems and scales efficiently to larger sizes and higher-band truncations far beyond the reach of exact diagonalization. We demonstrate the power of NTB in capturing diverse correlated phases in twisted MoTe$_2$, including charge density waves, fractional Chern insulators, and anomalous Hall Fermi liquids, within a unified framework. This approach offers a generic, scalable route towards understanding and discovering quantum phases in strongly correlated materials.
\end{abstract}

\maketitle
\blfootnote{*Email: ~\url{diluo1000@gmail.com}}

\textit{Introduction---.}
Recent advances in strongly correlated materials has lead to numerous experimental discoveries of unconventional physics, such as fractional quantum Hall (FQH) states \cite{zhang2020flat, lin2018moire}, superconductivity \cite{xia2024unconventional, xia2025superconductivity}, and generalized Wigner crystals \cite{regan2020mott, li2021imaging}. In particular, moiré superlattices in twisted transition-metal dichalcogenide homobilayers (tTMDs) generate strong electron correlations from flat band structures \cite{zhang2019nearly}, leading to novel states with nontrivial topology \cite{reddy2023toward, wang2024fractional, xu2023observation, kang2024observation}. External tuning of parameters can drastically alter their behavior \cite{liao2020precise, morales2021metal}, establishing tTMDs as a versatile platform for exploring correlated and topological quantum phenomena \cite{kennes2021moire, tang2020simulation}.

To theoretically describe strongly correlated materials, single particle Hamiltonian is obtained from ab initio calculations, such as the density functional theory (DFT). A fully quantum, many-body Hamiltonian including Coulomb interaction is then constructed from the single particle level of theory to correctly encode the correlations. To restrict the combinatorial growth of the Hilbert space while still preserving the essential physics, the fully quantum Hamiltonian is usually projected onto the lowest few bands, resulting in a multi-band projected (MBP) Hamiltonian \cite{reddy2023fractional, repellin2020ferromagnetism}. With limited band truncation for small systems, such Hamiltonian can be diagonalized exactly \cite{morales2021metal, li2021spontaneous, yu2024fractional,potasz2021exact, xie2021twisted, repellin2020ferromagnetism, carr2020electronic}. However, the resulting physics is often inaccurately captured due to strong band mixing and finite-size effects. In addition, mean field methods such as Hartree-Fock (HF) also fails to capture accurate physics due to long range interactions.

Meanwhile, recent advances in neural networks (NN) have demonstrated an alternative avenue to address such problem. With its remarkable representational power across all domains \cite{jumper2021highly, he2016deep, brown2020language, robledo2022fermionic}, NN has emerged as a compelling candidate for solving many-body problems. A natural framework for applying NN in this context is the neural quantum state (NQS) ~\cite{chen2023autoregressive,robledo2022fermionic, chen2022simulating, doi:10.1126/science.aag2302, Hibat_Allah_2020, PhysRevLett.124.020503, Irikura_2020, PhysRevResearch.3.023095, Han_2020,ferminet,Choo_2019,rnn_wavefunction,paulinet,Glasser_2018,Stokes_2020,Nomura_2017,martyn2022variational,Luo_2019,PhysRevLett.127.276402, https://doi.org/10.48550/arxiv.2101.07243,luo2022gauge, ma2025transformer}. 
For fermionic systems, neural network backflow (NNBF) has been proposed to encode fermionic statistics and strong correlation \cite{luo2019backflow}, inspiring later advancement for solving fermionic model Hamiltonians \cite{robledo2022fermionic, chen2025neural, gu2025solving} and quantum chemistry \cite{liu2024neural, liu2025efficient, pfau2020ab, hermann2020deep, von2022self}.

While it has been shown that NQS can accurately capture many-body physics under the real-space formalism ~\cite{cassella2023discovering,pescia2023message,kim2023neural,lou2023neural,entwistle2023electronic,wilson2022wave,li2022ab,scherbela2022solving,adams2021variational,smith2024ground,luo2023pairing,teng2024solving, qian2024taming,li2024emergent,Luo2024NNMoire, geier2025attention,romero2025spectroscopy}, incorporating NQS into the MBP formalism remains unexplored. In particular, despite the full Hilbert space without band truncation can be access in real space, enforcing momentum conservation becomes computationally expensive due to the need for symmetrization over all lattice translations. This makes the calculation of the energy spectrum of momentum resolved ground states, which contains important features such as degeneracies and energy gaps, computationally expensive. In contrast, in the MBP formalism, distinct momentum sectors are naturally decoupled due to momentum conservation.

In this work, we develop a neural Transformer backflow (NTB) framework, to solve the MBP Hamiltonian. Our architecture is mean-field transcendental, parameter-efficient, and fermionic intrinsic, exhibits superior performance compared to existing neural ansatzes. We also introduce a new sampling technique, which allows the calculation of the momentum-resolved many-body ground-state spectrum, revealing important information on energy degeneracy and gaps. By employing NTB to solve the MBP Hamiltonian for moir\'e materials,  we demonstrate that NTB not only agrees with exact diagonalization (ED) results for small systems with a few bands, but also extends beyond the limitations of ED, enabling access to higher bands and larger system sizes. Furthermore, we also show that NTB can accurately capture intricate quantum phases over various of filling factors, including charge density waves (CDW), fractional Chern insulator (FCI) and anomalous Hall Fermi liquids (AHFL), in moir\'e materials across a wide range of parameter regimes. This offers a unified approach for the simulation of strongly correlated materials within the MBP framework.

\textit{Multi-Band Projected Hamiltonian---.}
We start with the Hamiltonian of a generic many-body system defined on a periodic lattice. For a system that includes $N_k$ $\boldsymbol{k}$-points truncated at $N_b$ Bloch bands, the MBP Hamiltonian take the form of:
\begin{eqnarray}\label{BP_ham}
{\hat H}_{\rm MBP} &=& \sum_{i}^{N_s} \epsilon_{\boldsymbol{k}_i, n_i}c^\dagger_{\boldsymbol{k}_i, n_i} c_{\boldsymbol{k}_i, n_i} \nonumber  \\ 
&+& \frac{1}{2} \sum_{i,j,k,l}^{N_s^4} {\hat V}_{i,j,k,l} c^\dagger_{\boldsymbol{k}_i, n_i}c^\dagger_{\boldsymbol{k}_j, n_j} c_{\boldsymbol{k}_l, n_l}c_{\boldsymbol{k}_k, n_k}
\end{eqnarray}
where $N_s \equiv N_k N_b$ is the total number of single-particle Bloch states. $c^\dagger_{\boldsymbol{k}_i, n_i}$ creates a particle with momentum $\boldsymbol{k}_i$ in the $n_i^{th}$ Bloch band, and $\epsilon_{\boldsymbol{k}_i, n_i}$ originates from the dispersion relationship of the Bloch bands. Importantly, unlike orbitals in quantum chemistry, the overlaps between Bloch states are highly non-local due to periodicity of the system. This is reflected by the structure of Coulomb tensor, which takes the form of:
\begin{eqnarray}\label{coulomb_tensor}
{\hat V}_{i,j,k,l} = \frac{1}{A}\sum_{\boldsymbol q}\frac{2\pi e^2}{\epsilon |{\boldsymbol q}|}\bra{{{\boldsymbol k}_i}}e^{i{\boldsymbol q}{\boldsymbol r}_{1}}\ket{{{\boldsymbol k}_k}} \bra{{{\boldsymbol k}_j}}e^{-i{\boldsymbol q}{\boldsymbol r}_{2}}\ket{{{\boldsymbol k}_l}}
\end{eqnarray}
where $A$ is the supercell area. ${\boldsymbol q}$ represents momentum transfer between Bloch states. For arbitrary ${\boldsymbol q}$, it can always be decomposed as as a mesh component $[{\boldsymbol q}]$ that lives in the first Brillouin zone (FBZ), and an integer multiply of the reciprocal lattice (RL) vector $\boldsymbol{g}_{{\boldsymbol q}} \equiv {\boldsymbol q} - [{\boldsymbol q}]$. ${\hat V}_{i,j,k,l}$ only requires conservation of mesh component even if the RL component is not conserved. This makes ${\hat V}_{i,j,k,l}$ significantly long ranged and cannot be approximated by single determinant methods.

To examine the performance of NTB, we solve the MBP Hamiltonian for fractional moir\'e system. We start from the spin-$\uparrow$ component of the moir\'e continuum model \cite{wu2019topological}. In layer space, its form can be written as:
\begin{eqnarray}\label{continuous_hamil1}
\hat{H}_{\uparrow} = 
\begin{bmatrix}
\frac{-\hbar^2(\boldsymbol{k}-\boldsymbol{\kappa}_{+})^2}{2m^\ast} + V_{\mathfrak{b}}({\bf r})& t({\bf r}) \\
t^\dagger({\bf r}) & \frac{-\hbar^2(\boldsymbol{k}-\boldsymbol{\kappa}_{-})^2}{2m^\ast} + V_{\mathfrak{t}}({\bf r})
\end{bmatrix}~~~
\end{eqnarray}
Due to interlayer twisting, the $K$-points of the top and bottom layers are displaced by $\boldsymbol{\kappa}_{\pm}$. We choose the moir\'e reciprocal lattice vector as ${\bf g_i} = \frac{4\pi}{\sqrt{3}a_M}[\rm{cos}(\frac{\pi(i-1)}{3}){\hat x} + \rm{sin}(\frac{\pi(i-1)}{3}){\hat y}]$ with $\boldsymbol{\kappa}_{+} =\frac{ ({\bf g}_1 + {\bf g}_2)}{3}$ and $\boldsymbol{\kappa}_{-} = \frac{({\bf g}_1 + {\bf g}_6)}{3}$. $a_M = \frac{a_0}{2\rm{sin}(\theta /2)}$, where $a_0$ is the monolayer lattice constant. The periodic moir\'e potential $V_{l}$ ($l = \mathfrak{b}, \mathfrak{t}$) takes the form of:
\begin{eqnarray}
V_{l}({\bf r}) = -2V \sum_{i=1,3,5}\rm{cos}({\bf g_i \cdot r} + \phi_{l} )
\end{eqnarray}
 where $\phi_{\mathfrak{b}} = -\phi_{\mathfrak{t}} = \phi$. The interlayer tunneling amplitude $t({\bf r})$ takes the form of \cite{bistritzer2011moire}:
\begin{eqnarray}
t({\bf r}) = \omega(1+e^{i{\bf g}_2 {\bf r}} + e^{i{\bf g}_3 {\bf r}})
\end{eqnarray}
By numerically diagonalizing Eq.~\ref{continuous_hamil1}, we obtain $\epsilon_{\boldsymbol{k}_i, n_i}$, as well as the Bloch state coefficients (details can be found in \cite{supmat}). For the rest of this work, we adopt the following parameters for tMoTe$_2$ \cite{reddy2023fractional}: $(V, \omega, \phi, m^\ast, \theta)$ = ($11.2$ meV, $13.3$ meV, $-91 ^\circ$, 0.62 $m_e$, $2.7^\circ$).

\begin{figure}[t]
\centering
\centerline{\includegraphics[width=80mm]{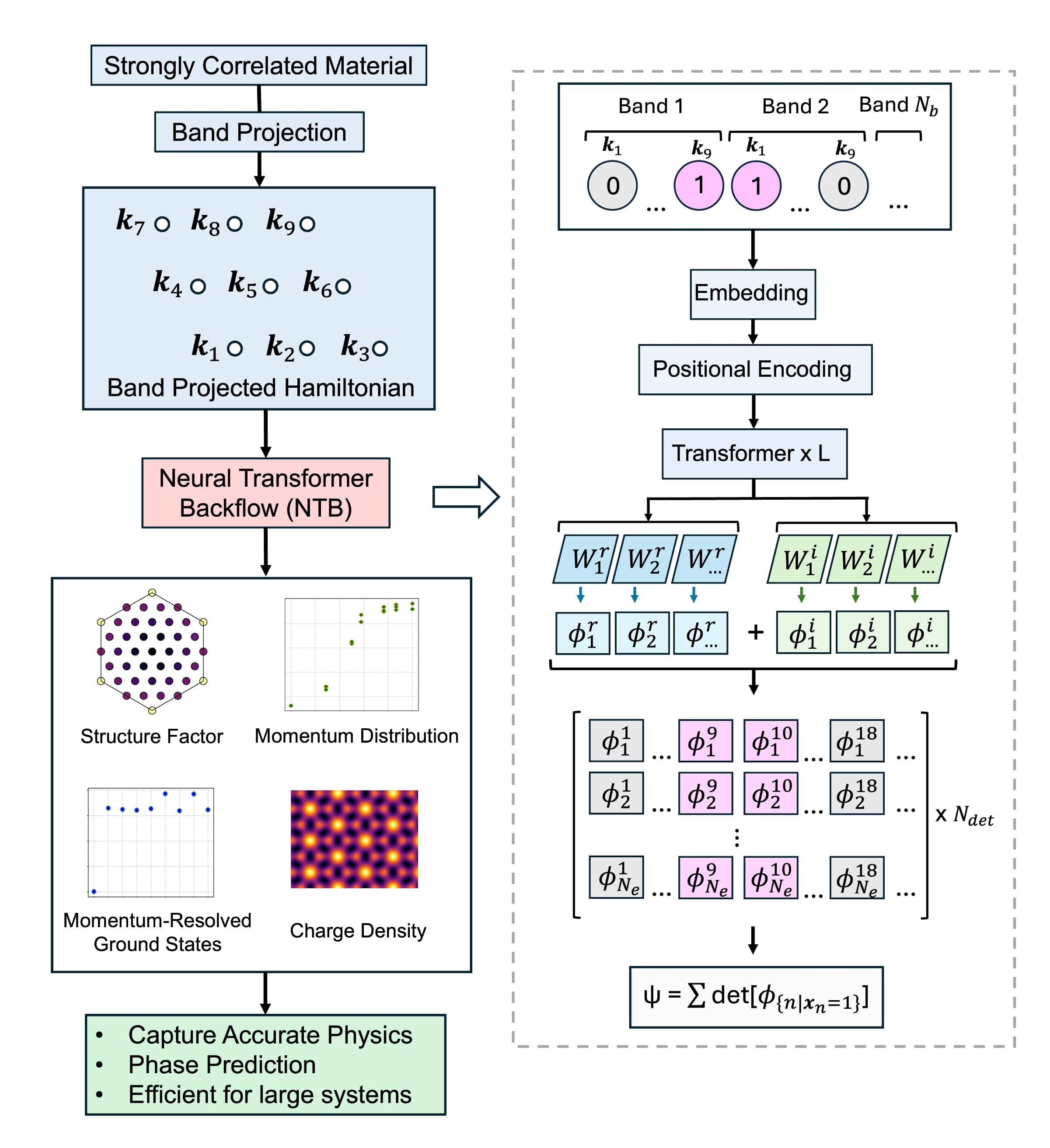}}
\caption{Schematic of the architecture and applications of the Neural Transformer Backflow (NTB).}
\label{flowchart}
\end{figure}

\textit{Neural Transformer Backflow---.}
Notably, the Hilbert space of ${\hat H}_{\rm MBP}$ grows exponentially, which makes ED impractical for large system sizes. To address this issue, we employ NTB to solve MBP Hamiltonian variationally. The general form of NTB ansatz can be expressed as:
\begin{eqnarray}\label{NQS}
\ket{\Psi_\theta} = \sum_{{\boldsymbol x}} \psi_\theta({\boldsymbol x})\ket{{\boldsymbol x}}
\end{eqnarray}
where $\ket{\boldsymbol x} = \ket{x_1, x_2, \dots, x_n}$, with $x_n \in \{0,1\}$, denotes the basis states of $\hat{H}_{\rm{MBP}}$. Therefore, $n = 1,2,\dots,N_s$. The amplitude $\psi_\theta({\boldsymbol x})$ is obtained by summing the determinants of neural orbitals:
\begin{eqnarray}\label{sumdet}
\psi_\theta({\boldsymbol x}) = \sum_{k=1}^{N_{\rm det}}\text{det}[\phi^{k}_{\{n|x_n = 1\}}({\boldsymbol x}, \theta)]
\end{eqnarray}
where $N_{\rm det}$ is the number of determinants. We note that the neural orbitals depends on all $x_n$, which is known as the backflow transformation. The neural orbitals are selected based on the occupation pattern of ${\boldsymbol x}$ ($\phi^{k}_{\{n|x_n = 1\}}$ indicates that neural orbitals with $x_n = 0$ are excluded from the determinants). To generate $\phi_n^k$, ${\boldsymbol x}$ is first embedded with positional encoding, and then input into a multi-head Transformer. For the $n^{th}$ orbital, the Transformer output can be written as ${\bf h}_n = {\rm Transformer}({\boldsymbol x})_n$. Per-orbital linear layers are used to transform ${\bf h}_n$ into $N_e \times N_{\rm det}$ neural orbitals via: ${\Phi}_n({\boldsymbol x}, \theta) = [{W}_n^{r}({{\bf h}_n}) + i{W}_n^{i}({{\bf h}_n}) + (b_n^{\rm r} + i b_n^{\rm i})]$, where ${\Phi}_n \in \mathbb{C}^{N_{\rm det}\times N_e}$. ${\phi}_n^k$ is the $k^{\rm th}$ determinant component of ${\Phi}_n$, i.e. ${\Phi}_n = [{\phi}_n^1, {\phi}_n^2, \dots, {\phi}_n^{N_{\rm det}}]$.

\begin{figure}[t]
\centering
\centerline{\includegraphics[width=90mm]{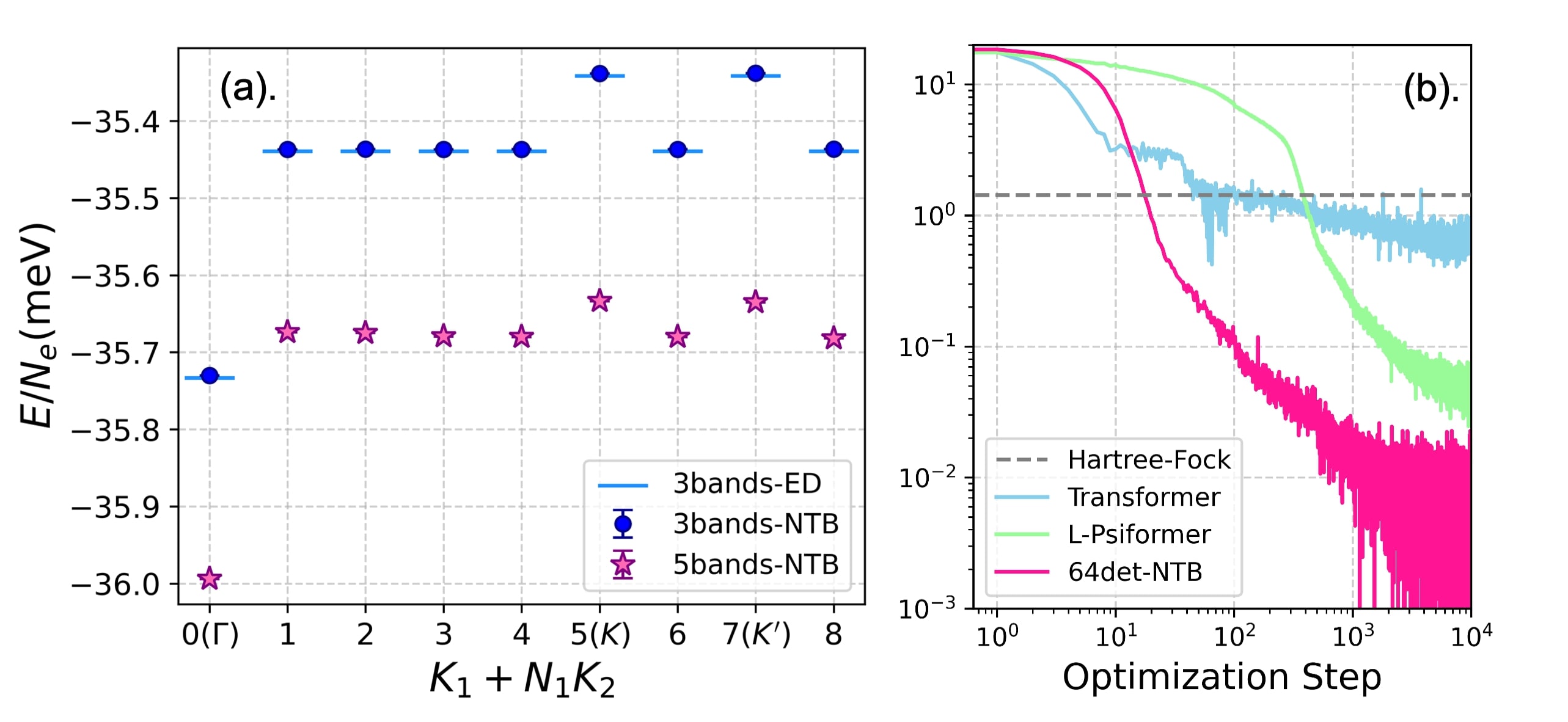}}
\caption{(a). Energy spectrum of momentum resolved ground states across different momentum sectors. 3-bands ED results are plotted against NTB results with 3-bands and 5-bands. (b). Optimization curve of $\Gamma$-point energy with $N_b = 3$ for different NN architectures. The gray dotted line is reference energy obtained from HF method. All networks are trained with $N_{\rm batch} = 10^{4}$. For (a) and (b), the parameters used are: $N_{\rm site} = 9$, $\nu = 2/3$, $\epsilon/\epsilon_0 = 10$, $\theta = 2.7^{\circ}$.}
\label{9site_spec}
\end{figure}

The advancement and novelty in NTB architecture stem from three important properties:

\begin{enumerate}[leftmargin=*]
\item{\textit{Mean-field transcendental.}~}By simply setting $N_{\rm hid} = N_{\rm det} = 1$ and ${\bf h}_n = x_n$, ${W}_n$ corresponds exactly to the mean field orbitals, reducing the architecture to the same mathematical structure as HF method. This allows NTB to naturally encodes many-body effects into the mean-field framework. Powered by Transformer's  expressivity, NTB is able to capture many-body correlations, transcending the performance of mean-field methods.
\item{\textit{Parameter-efficient.}~}The linear layers are designed per orbital, which avoids a fully dense connection between the Transformer output and neural orbitals. This substantially decrease the total number of network parameter, allowing further augmentation of network performance by increasing $N_{\rm det}$. 
\item{\textit{Fermionic intrinsic.}~}The final amplitude is obtained by taking configuration-dependent Slater determinants, which preserved fermionic statistics intrinsically, different from the previous work on direct application of Transformer.
\end{enumerate}
For the rest of this paper, we use a hidden dimension $N_{\rm hid} = 32$, number of layers $N_{\rm layer} = 2$ and number of attention heads $N_{\rm head}=16$.

The network parameters are then updated following the variational Monte-Carlo(VMC) method. Under the VMC framework, the variational energy can be written as:
\begin{eqnarray}\label{NQS_energy}
E_\theta = \frac{\bra{\Psi_\theta}{\hat H}_{\rm MBP}\ket{\Psi_\theta}}{\langle \Psi_\theta|\Psi_\theta\rangle } = \mathbb{E}_{{\boldsymbol x} \sim |\psi_\theta({\boldsymbol x})|^2}[E_{\rm loc}({\boldsymbol x})]
\end{eqnarray}
where $E_{\rm loc}({\boldsymbol x}) = \frac{\sum_{{\boldsymbol x^\prime}} h_{{\boldsymbol x^\prime}, {\boldsymbol x}} \psi_\theta({\boldsymbol x^\prime})}{\psi_\theta({\boldsymbol x})}$ is the local energy, $h_{{\boldsymbol x^\prime}, {\boldsymbol x}} = \bra{{\boldsymbol x^\prime}}{\hat H}_{\rm MBP} \ket{{\boldsymbol x}}$ is the matrix element connecting ${\boldsymbol x^\prime}$ and ${\boldsymbol x}$. The gradient of $E_\theta$ is then given by:
\begin{eqnarray}\label{Energy_gradient}
\nabla_{\theta} E_\theta = 2{\rm Re}\Big\{\mathbb{E}_{{\boldsymbol x} \sim |\psi_\theta({\boldsymbol x})|^2}\Big[\frac{\partial~{\rm ln}|\psi_\theta({\boldsymbol x})|}{\partial\theta}[E_{\rm loc}({\boldsymbol x}) - E_\theta]\Big]\Big\} \nonumber \\
\end{eqnarray}
By sampling ${\boldsymbol x}$ from $|\psi_\theta({\boldsymbol x})|^2$, $\nabla_{\theta} E_\theta$ can be approximated . To avoid local minimum, we use the signSGD method \cite{bernstein2018signsgd} to stochastically update $\theta$. We then employ the Markovian-chain Monte-Carlo(MCMC) method to sample $N_{\rm batch}$ basis states for every optimization step.

\begin{figure}[t]
\centering
\centerline{\includegraphics[width=80mm]{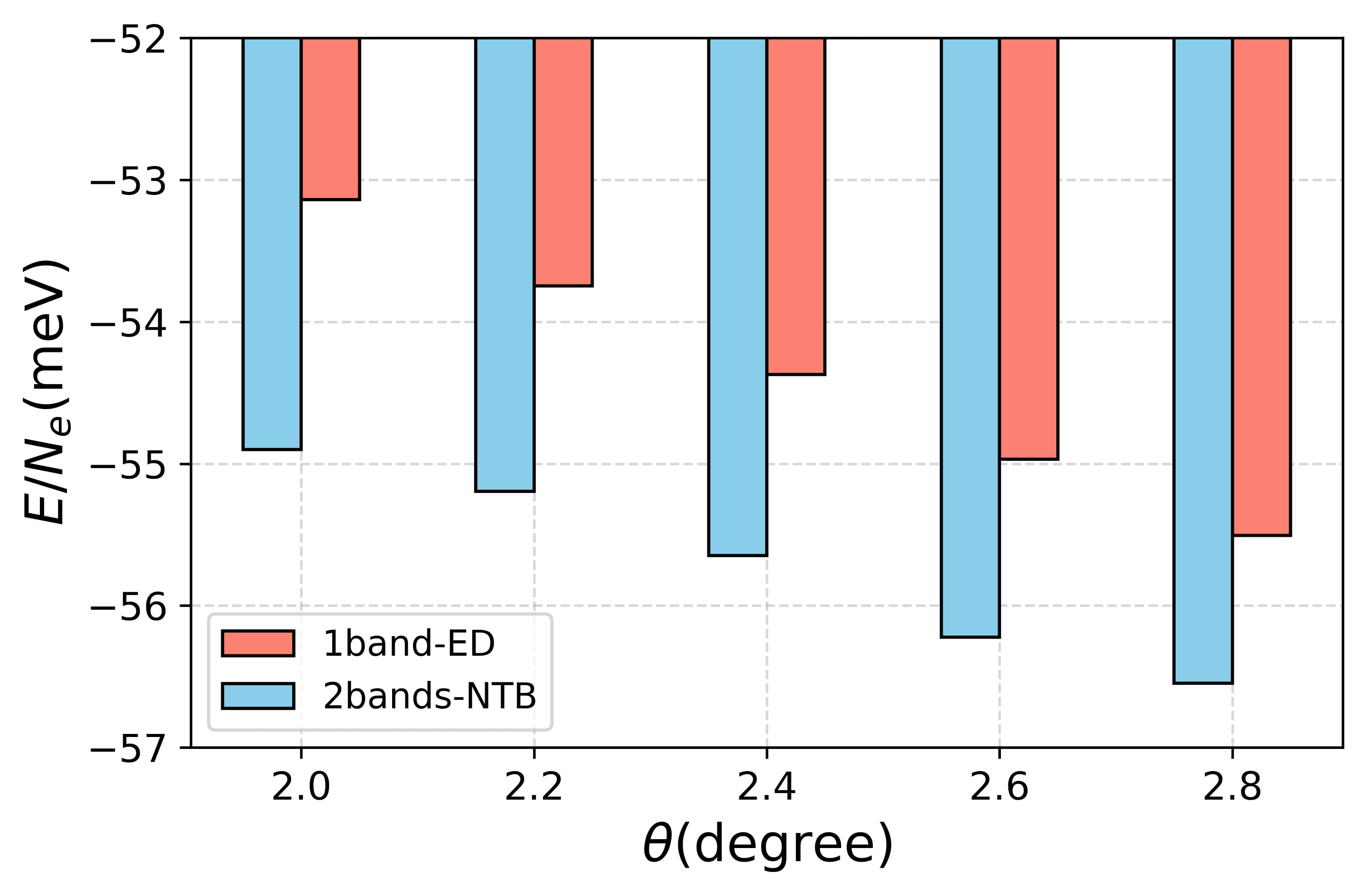}}
\caption{$\Gamma$-point energy across different choices of $\theta$. 1-band ED results are plotted against 2-bands NTB results. The parameters used are: $N_{\rm site} = 25$, $\nu = 3/5$, $\epsilon/\epsilon_0 = 5$.}
\label{25site_eps}
\end{figure}

Notably, NTB is designed to automatically conserve total momentum, which enables the calculation of momentum-resolved ground states. This is achieved by introducing an advanced MCMC technique with momentum conservation. For a sample ${\boldsymbol x}$ with $x_i =x_j = 0$ and $x_k = x_l = 1$, we define a momentum conserved flip set:
\begin{eqnarray}
S &=& \{ (i,j;k,l) | [\boldsymbol{k}_i + \boldsymbol{k}_j] = [\boldsymbol{k}_k + \boldsymbol{k}_l] \}
\end{eqnarray}
By drawning proposals randomly from $S$ and accepted based on the Metropolis–Hastings algorithm, we are able to restrict the sampling space within a symmetry sector of the Hilbert space with fixed center-of-mass (CM) momentum. Within such a sector, the number of non-zero $h_{{\boldsymbol x^\prime}, {\boldsymbol x}}$ for a given ${\boldsymbol x}$ scales as $\mathcal{O}(N_e^2 N_s) \approx \mathcal{O}(N_e^3)$, which remains tractable even for large system sizes. We also use GPU accelerations with Pytorch \cite{paszke2019pytorch} to further boost the computational speed. 

Moreover, we note that the inference time with momentum conservation scales with $\mathcal{O}(N_e^3)$ for our method, as momentum is naturally conserved during sampling. This is in contrast with the $\mathcal{O}(N_e^4)$ scaling for wavefunction symmetrization approaches \cite{li2025deep}, where the the extra factor arises from averaging over lattice translations. We also note that transfer learning can be naturally implemented across different momentum sectors, which significantly reduces the optimization time.

\textit{Results---.} We first benchmark ground state energies computed by NTB on a $3 \times 3$ cluster with ED results. An arbitrary momentum on lattice can be expressed as ${\boldsymbol k} = K_1 {\boldsymbol T}_1 + K_2 {\boldsymbol T}_2$, where ${\boldsymbol T}_i$ is the basis vectors of reciprocal lattice. We assign an integer index $K_1 + N_{1} K_2$ to every unique allowed momentum in the FBZ ($N_1 \times N_2 = N_{\rm site}$). In Fig.~\ref{9site_spec}a, we set $\epsilon = 10$ and $\nu = 2/3$. With $N_b = 3$, the NTB results show excellent agreement with ED. When $N_b = 5$, NTB generates much lower energies across all momentum sectors comparing to ED with $N_b = 3$. It is also clear that NTB could capture the symmetry of the system by noting energy degeneracies across different momentum sectors. In Fig.~\ref{9site_spec}b, we benchmark NTB with $2$ different architectures on the $\Gamma$-point energy of the same system. The first is Lattice-Psiformer(L-Psiformer), which is a lattice variant of Ref.~\cite{von2022self}. The second is a simple Transformer that evaluates $\psi_{\theta}$ auto-regressively(Details on these architectures can be find in  \cite{supmat}). The number of parameters is kept at approximately $700$k across all three architectures for fairness. Although all architectures achieve a lower energy than HF (HF energy is calculated using the algorithm from Ref.~\cite{dai2024strong}), NTB outperforms other methods by achieving the lowest energy with a deviations of only $3 \times 10^{-3}$ meV.

\begin{figure}[t]
\centering
\centerline{\includegraphics[width=80mm]{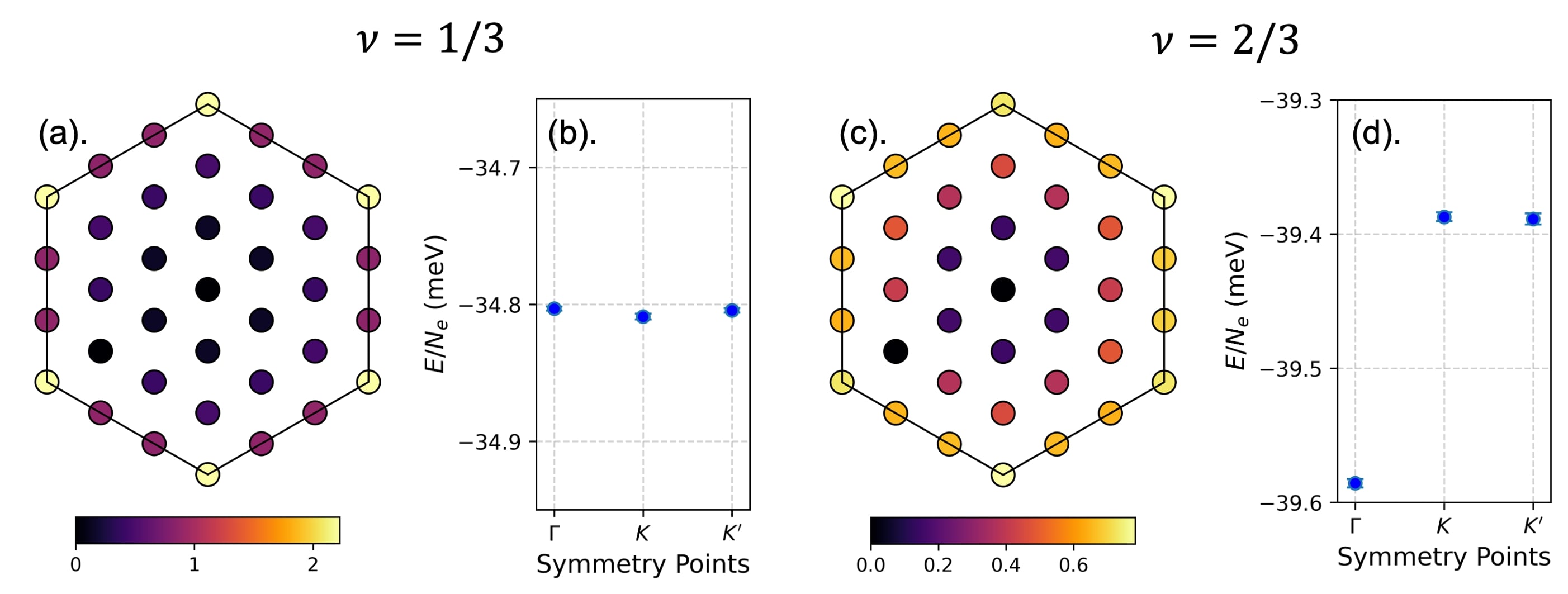}}
\caption{Energy spectrum of momentum resolved ground states and the corresponding structure factor for filling factors $\nu=1/3$ and $\nu=2/3$. (a), (c): Structure factor calculated with $\Gamma$-point ground state. (b), (e): Energy spectrum of momentum resolved ground states at high symmetry points. The parameters used are: $N_{\rm site} = 27$, $N_b = 2$, $\epsilon/\epsilon_0 = 10$, $\theta = 2.7^{\circ}$. }
\label{27site_fig1}
\end{figure}

In Fig.~\ref{25site_eps}, we present the ground state energy at $\Gamma$-point for twist angles $\theta = 2.0^\circ$, $2.2^\circ$, $2.4^\circ$, $2.6^\circ$, and $2.8^\circ$ on a $5 \times 5$ cluster at filling $\nu = 3/5$. Leveraging the $\mathcal{O}(N_e^3)$ scaling of the NTB architecture, we are able to sample from ${\hat H}_{\rm MBP}$ with $N_b = 2$. The corresponding Hilbert space size is $2.25 \times 10^{12}$, which far exceeds the maximum capacity of ED. We find that NTB achieves per-particle energies around 1.5 meV lower than those of ED with $N_b = 1$, highlighting NTB’s superior scalability and efficiency far beyond the reach of ED. Moreover, the monotonic decrease of NTB energies with $\theta$ also indicates that NTB captures band mixing across different angles accurately.

In Fig.~\ref{27site_fig1}, we plot the energy spectrum of momentum-resolved ground states at high symmetry points ($\Gamma$, $K$, $K^\prime$) and the $\Gamma$ point structure factor at filling factors $\nu=1/3$ and $\nu=2/3$. The results are computed by NTB with $64$ determinants on a $N_{\rm site} = 27$ cluster (27A in Ref.~\cite{wilhelm2021interplay}) with $N_b = 2$, which also far beyond the limit of ED. We choose $\theta = 2.7^{^\circ}$, which is above the magic angle $\theta_m \approx 2.0^{^\circ}$ \cite{reddy2023fractional}. In Fig.~\ref{27site_fig1}a, at fillings $\nu=1/3$, we identify the emergence of CDW phase by observing sharp peaks in $S(\mathbf q)$ at $K$-points. Furthermore, in Fig.~\ref{27site_fig1}b, we find nearly degenerate states at the three high symmetry points, which further confirms with the characteristics of the CDW phase \cite{reddy2023toward}. For $\nu=2/3$, we observe no peak in $S(\mathbf q)$ plotted in Fig.~\ref{27site_fig1}c, and a sizable gap between $\Gamma$ and $K$, $K^\prime$ points, as shown in Fig.~\ref{27site_fig1}d. This agrees with the FCI phase. The FCI phase is known to have degenerate ground states that permute under flux insertion. For the $N_{\rm site} = 27$ cluster, only the $\Gamma$-point remains invariant under such transformations, which agrees with our results.

\begin{figure}[b]
\centering
\centerline{\includegraphics[width=90mm]{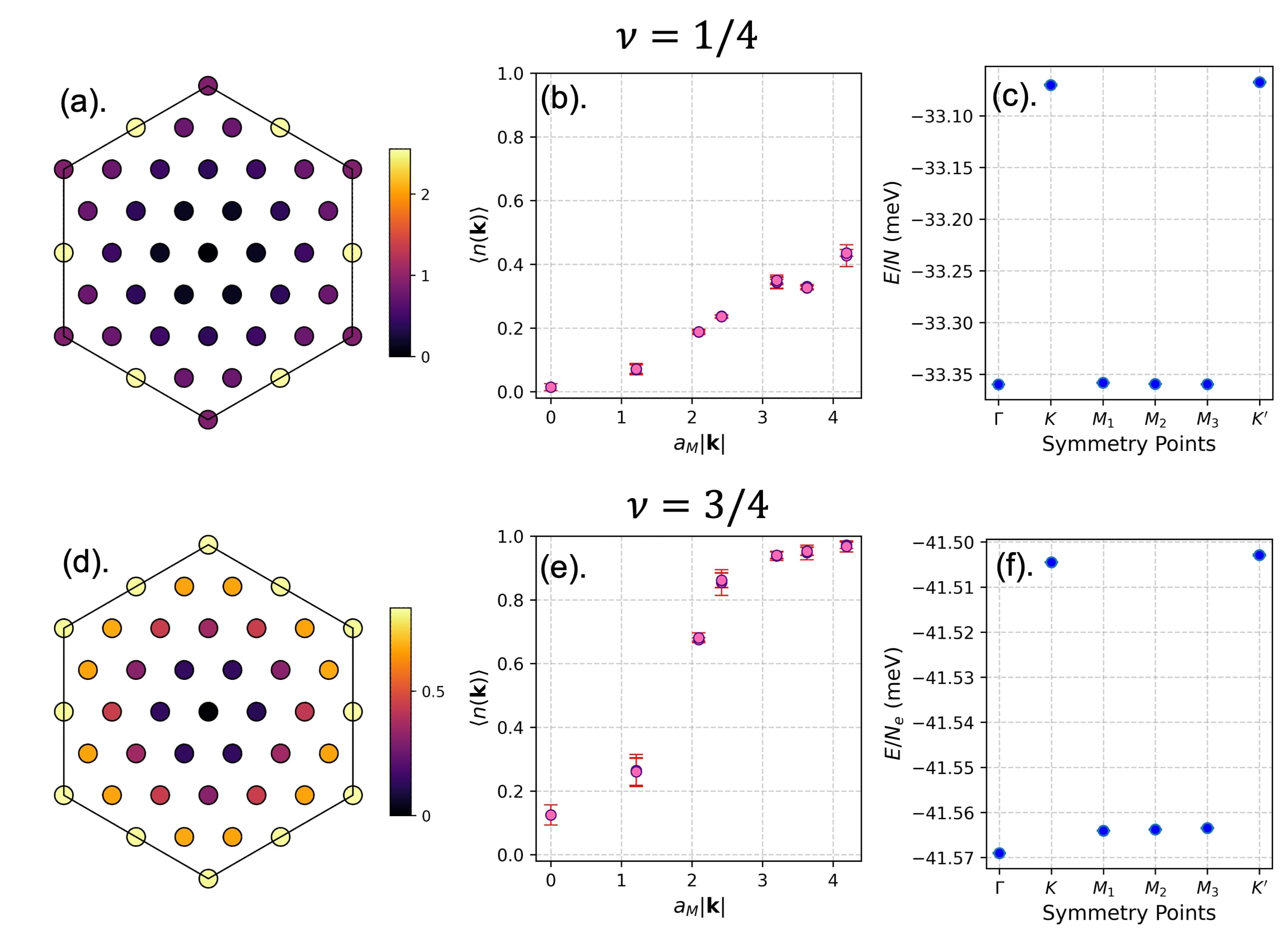}}
\caption{Various of observables for filling factors $\nu= 1/4$ and $\nu= 3/4$. (a), (d): Structure factor at $\Gamma$-point. (b), (e): Momentum distribution function at $\Gamma$-point. (c), (f): Energy spectrum of momentum resolved ground states at high symmetry points. The parameters used are: $N_{\rm site} = 36$, $N_b = 1$, $\epsilon/\epsilon_0 = 10$, $\theta = 2.7^{\circ}$.}
\label{36site_fig2}
\end{figure}

In Fig.~\ref{36site_fig2}, we plot various of observables at filling factors $\nu = 1/4$ and $\nu = 3/4$ calculated on a $6 \times 6$ cluster with $N_b = 1$, which is also challenging for ED. For $\nu = 1/4$, we observe a sharp peak in $S(\mathbf q)$ at $M$-points, which indicates a different CDW phase compared with Fig.~\ref{27site_fig1}a. Moreover, four degenerate ground states are observed in Fig.~\ref{36site_fig2}c, which agrees with the number of degenerate ground states for the CDW phase at $\nu = 1/4$. For $\nu = 3/4$, as no significant peak is observed in $S(\mathbf q)$ in Fig.~\ref{36site_fig2}d, a sharp Fermi surface around $a_M|\mathbf{k}| \sim 2$ is observed in Fig.~\ref{36site_fig2}e. This confirms the presence of the AHFL phase. Moreover, due to the formation of quasiparticles in AHFL phase, the ground state degeneracies can be explained by filling electrons into the mean-field orbitals. At filling $\nu = 3/4$ with $N_e = 27$, the lowest 23 electrons occupy the lowest-energy orbitals, leaving 6 degenerate orbitals for the remaining 4 electrons. This yields $C_4^6 = 15$ degenerate many-body states, distributed across 10 distinct momentum sectors—including $\Gamma$ and $M$ points, but not $K$ points \cite{supmat}. This confirms with the energy spectrum in Fig.~\ref{36site_fig2}f, where the $\Gamma$ and $M$ points are quasi-degenerate.

\textit{Discussions---.} In this work, we have introduced a new neural-network backflow framework to solve the MBP Hamiltonians of strongly correlated materials. NTB is mean-field transcendental, parameter-efficient, and fermionic intrinsic, demonstrating excelling performances compared to existing architectures. Moreover, it goes beyond the limit of ED by including higher bands and larger system sizes, achieving a much lower ground state energy. By enforcing momentum conservation during sampling, our method allows efficient calculation of momentum-resolved ground states across momentum sectors, providing a more efficient tool towards a more comprehensive understanding of strongly correlated materials.  In addition, by  calculating observables such as $S({\bf{q}})$ and $n({\bf{k}})$, we demonstrate that NTB can capture a wide range of nontrivial topological and correlated quantum states. Future directions include exploring the calculation of superconductivity and topological quantities, such as the Chern number, within the band projected framework. Leveraging NTB's scalability and the ability to probe a wide range of observables, our approach offers the potential to map out a global phase diagram for various quantum materials across different parameter regimes.

\textit{Acknowledgement}.---The authors thank Aidan P. Reddy and Timothy Zaklama for useful discussions.

\bibliography{main}

\bibliographystyle{apsrev4-1}

\appendix

\clearpage

\onecolumngrid
\begin{center}
	\noindent\textbf{Supplementary Material}
	\bigskip
		
	\noindent\textbf{\large{}}
\end{center}

\onecolumngrid

\section{Details on Band Projected Hamiltonian}
We start from the spin-$\uparrow$ component of the continuum model \cite{wu2019topological}. In layer space, its form can be written as:
\begin{eqnarray}\label{continuous_hamil}
\hat{H}_{\uparrow} = 
\begin{bmatrix}
\frac{-\hbar^2(\nabla-\boldsymbol{\kappa}_{+})^2}{2m^\ast} + V_{\mathfrak{b}}({\bf r})& t({\bf r}) \\
t^\dagger({\bf r}) & \frac{-\hbar^2(\nabla-\boldsymbol{\kappa}_{-})^2}{2m^\ast} + V_{\mathfrak{t}}({\bf r})
\end{bmatrix}~~~
\end{eqnarray}
Due to interlayer twisting, the $K$-points of the top($\mathfrak{t}$) and bottom($\mathfrak{b}$) layers are displaced by $\boldsymbol{\kappa}_{\pm}$. We choose the moir\'e reciprocal lattice vector as ${\bf g_i} = \frac{4\pi}{\sqrt{3}a_M}[\rm{cos}(\frac{\pi(i-1)}{3}){\hat x} + \rm{sin}(\frac{\pi(i-1)}{3}){\hat y}]$ ($i \le 6$) with $\boldsymbol{\kappa}_{+} =\frac{ ({\bf g}_1 + {\bf g}_2)}{3}$ and $\boldsymbol{\kappa}_{-} = \frac{({\bf g}_1 + {\bf g}_6)}{3}$. $a_M = \frac{a_0}{2\rm{sin}(\theta /2)}$, where $a_0$ is the monolayer lattice constant. The periodic moir\'e potential $V_{l}$ ($l = \mathfrak{b}, \mathfrak{t}$) takes the form of:
\begin{eqnarray}
V_{l}({\bf r}) = -2V \sum_{i=1,3,5}\rm{cos}({\bf g_i \cdot r} + \phi_{l} )
\end{eqnarray}
 where $\phi_{\mathfrak{b}} = -\phi_{\mathfrak{t}} = \phi$. The interlayer tunneling amplitude $t({\bf r})$ takes the form of \cite{bistritzer2011moire}:
\begin{eqnarray}
t({\bf r}) = \omega(1+e^{i{\bf g}_2 {\bf r}} + e^{i{\bf g}_3 {\bf r}})
\end{eqnarray}
In this work, we assume full spin polarization, and the single particle Hamiltonian in second quantization is ${\hat H}_0 = \int d\boldsymbol{r} \psi^\dagger(\boldsymbol{r}) \hat{H}_{\uparrow} \psi(\boldsymbol{r})$. For arbitrary momentum in reciprocal space, its mesh component can be written as $[\boldsymbol{q}] = \boldsymbol{q} - (m_{1} \boldsymbol{b}_1 + m_{2} \boldsymbol{b}_2)
$ ($m_{i} \in \mathbb{Z}$), where $\boldsymbol{b}_1$ and $\boldsymbol{b}_2$ are the basis of reciprocal lattice (We choose ${\boldsymbol{b}_1} = \boldsymbol{g}_1$ and ${\boldsymbol{b}_2} = \boldsymbol{g}_3$ from here onwards). Similarly, the reciprocal lattice component of $\boldsymbol{q}$ can be written as $\boldsymbol{g}_{\boldsymbol{q}} = m_{1} \boldsymbol{b}_1 + m_{2} \boldsymbol{b}_2$. We denote $\boldsymbol{k}$ as an arbitrary momentum in the First Brillouin Zone (FBZ), and $\boldsymbol{g}$ as an arbitrary reciprocal lattice vector. Fourier expanding $\psi^\dagger(\boldsymbol{r})$ and $\psi^(\boldsymbol{r})$ enables diagonalization of ${\hat H}_0$ in $\boldsymbol{k} + \boldsymbol{g}$ basis, yielding Bloch states and band energies:
\begin{eqnarray}
\ket{u_{\boldsymbol{k}, n}} = \sum_{\boldsymbol{g}, l}u_{\boldsymbol{k},l}^{(n)}(\boldsymbol{g}) \ket{\boldsymbol{k} + \boldsymbol{g},l} ~~~~~\text{with}~~~~~ {\hat H}_{0}(\boldsymbol{k}) \ket{u_{\boldsymbol{k}, n}} = \epsilon_{\boldsymbol{k}, n} \ket{u_{\boldsymbol{k}, n}}
\end{eqnarray}
where $n$ is band index. To construct the many-body Hamiltonian, we define Bloch state creation and annihilation operator ${\hat c}_{\boldsymbol{k}, n}^\dagger$ and ${\hat c}_{\boldsymbol{k}, n}$. The many-body wavefunction is effectively projected onto a bitstring basis $\ket{\boldsymbol x} = \ket{x_1, x_2, \dots, x_{N_{q}} }$, where each $x_j \in \{0,1\}$ encodes the occupation of the Bloch state $\ket{u_{\boldsymbol{k}_j, n_j}}$. $N_{s} = N_{k}\times N_{b}$ is the total number of single-particle Bloch states, and $N_{b}$ is the total number of bands. Numerically, we restrict $\ket{u_{\boldsymbol{k}, n}}$ such that $ \epsilon_{\boldsymbol{k}, n} \in N_b \text{ lowest eigenvalues of }\hat{H}_{\uparrow}(\boldsymbol{k} )$. This leads to the following Multi-Band Projected(MBP) Hamiltonian:
\begin{eqnarray}\label{BP_ham}
{\hat H}_{\rm MBP} = \sum_{i} \epsilon_{\boldsymbol{k}_i, n_i}c^\dagger_{\boldsymbol{k}_i, n_i} c_{\boldsymbol{k}_i, n_i} + \frac{1}{2} \sum_{i,j,k,l} {\hat V}_{i,j,k,l} c^\dagger_{\boldsymbol{k}_i, n_i}c^\dagger_{\boldsymbol{k}_j, n_j} c_{\boldsymbol{k}_l, n_l}c_{\boldsymbol{k}_k, n_k}
\end{eqnarray}
where $i,j,k,l$ are bit indices. The Coulomb tensor takes the form of:
\begin{eqnarray}\label{coulomb_tensor}
{\hat V}_{i,j,k,l} &=& \frac{1}{A}\sum_{\boldsymbol q} V({\boldsymbol q})\bra{u_{{\boldsymbol k}_i, n_i}}e^{i{\boldsymbol q}{\boldsymbol r}_{1}}\ket{u_{{\boldsymbol k}_k, n_k}} \bra{u_{{\boldsymbol k}_j,n_j}}e^{-i{\boldsymbol q}{\boldsymbol r}_{2}}\ket{u_{{\boldsymbol k}_l,n_l}} \nonumber \\
&=& \frac{1}{A} \sum_{\boldsymbol q} V({\boldsymbol q}) \delta_{{\boldsymbol k}_i, [{\boldsymbol k}_k + {\boldsymbol q}]} \delta_{{\boldsymbol k}_j, [{\boldsymbol k}_l - {\boldsymbol q}]}F({\boldsymbol k}_i, {\boldsymbol k}_k , n_i, n_k, {\boldsymbol g}_{{\boldsymbol k}_k + {\boldsymbol q}}) F({\boldsymbol k}_j, {\boldsymbol k}_l, n_j, n_l, {\boldsymbol g}_{{\boldsymbol k}_l - {\boldsymbol q}})
\end{eqnarray}
where $V({\boldsymbol q}) = \frac{2\pi e^2}{\epsilon |{\boldsymbol q}|}$. $A = N_{k} \cdot|{\boldsymbol b}_1 \times {\boldsymbol b}_2|$ is the supercell area. $F(\boldsymbol{k}_1,\boldsymbol{k}_2, n_1, n_2, \boldsymbol{g}) = \sum_{\boldsymbol{g}^\prime, l}u_{\boldsymbol{k}_1,l}^{(n_1)\ast}(\boldsymbol{g}^\prime+ \boldsymbol{g})u_{\boldsymbol{k}_2,l}^{(n_2)}(\boldsymbol{g}^\prime) $ is the form factor. Numerically, ${\boldsymbol q}$ should be summed over all possible momentum transfer allowed the $\boldsymbol{k} + \boldsymbol{g}$ basis.
\\

We note that the Hilbert space of ${\hat H}_{\rm MBP}$ scales with $N_s\choose N_e$. Each bitstring is connected with $\mathcal{O}(N_e^2 N_s)$ others via two-body interactions. At small and intermediate system sizes, it is possible to construct ${\hat H}_{\rm MBP}$ as a sparse matrix and diagonalize it to get many-body ground state. However, pre-computing all non-zero elements of ${\hat H}_{\rm MBP}$ is computationally infeasible for large system size. Instead, we compute a fraction of ${\hat H}_{\rm MBP}$ on-the-fly to avoid memory overflow. Firstly, we compute and store ${\hat V}_{i,j,k,l}$. Due to momentum conservation, the number of non-zero elements in ${\hat V}_{i,j,k,l}$ scales with $\mathcal{O}(N_e^3N_b)$ instead of $\mathcal{O}(N_e^4)$. Nextly, we identify and store all $(\boldsymbol{k}_i, \boldsymbol{k}_j)$ such that their respective $[\boldsymbol{k}_i + \boldsymbol{k}_j]$ are equal. This allows us to locate all bitstrings in the connected space fastly.

\section{Details on Structure Factor}
In momentum space, the structure factor can be written as \cite{zaklama2024structure}:
\begin{eqnarray}\label{sup_Sq}
S({\boldsymbol q}) = \frac{1}{A}\langle \rho({\boldsymbol q})\rho(-{\boldsymbol q})\rangle
\end{eqnarray}
where $\rho({\boldsymbol q}) = \sum_{\boldsymbol{k}}{\hat f}_{\boldsymbol{k}}^\dagger {\hat f}_{\boldsymbol{k}+\boldsymbol{q}}$ is the generic form of fermionic density operator in momentum space. In the setting of moir\'e system, ${\hat f}_{\boldsymbol{k}}$ carries internal DOFs, including layer quantum number $l$ and reciprocal lattice vector $\boldsymbol{g}$.  Under band projection, the fermionic operators can be decomposed into:
\begin{eqnarray}\label{sup_foperators}
{\hat f}_{\boldsymbol{k}, l, \boldsymbol{g}} = \sum_{n} u_{\boldsymbol{k},l}^{(n)}(\boldsymbol{g}) {\hat c}_{\boldsymbol{k},n}
\end{eqnarray}
where $u_{\boldsymbol{k},l}^{(n)}(\boldsymbol{g})$ is the amplitude of Bloch wavefunction in $(\boldsymbol{k} + \boldsymbol{g})$ basis. By folding $\boldsymbol{k}+\boldsymbol{q}$ back to FBZ and summing up all internal DOFs, the density operator can be written as:
\begin{eqnarray}\label{sup_density}
\rho({\boldsymbol q}) = \sum_{\boldsymbol{k}, l, \boldsymbol{g}}{\hat f}_{\boldsymbol{k},l,\boldsymbol{g}}^\dagger {\hat f}_{\boldsymbol{k}+\boldsymbol{q}, l, \boldsymbol{g}} = \sum_{\boldsymbol{k}, l, \boldsymbol{g}}\sum_{n_1, n_2} u_{\boldsymbol{k},l}^{(n_1)\ast}(\boldsymbol{g})u_{[\boldsymbol{k}+\boldsymbol{q}],l}^{(n_2)}(\boldsymbol{g} + \boldsymbol{g}_{\boldsymbol{k}+\boldsymbol{q}}) {\hat c}_{\boldsymbol{k},n_1}^\dagger {\hat c}_{[\boldsymbol{k}+\boldsymbol{q}],n_2} 
\end{eqnarray}
Let $\boldsymbol{g}^\prime = \boldsymbol{g} + \boldsymbol{g}_{\boldsymbol{k}+\boldsymbol{q}}$, $\rho({\boldsymbol q})$ can be written in terms of form factors:
\begin{eqnarray}\label{sup_density2}
\rho({\boldsymbol q}) &=& \sum_{\boldsymbol{k}}\sum_{n_1, n_2}[\sum_{\boldsymbol{g}^\prime, l} u_{\boldsymbol{k},l}^{(n_1)\ast}(\boldsymbol{g}^\prime - \boldsymbol{g}_{\boldsymbol{k}+\boldsymbol{q}})u_{[\boldsymbol{k}+\boldsymbol{q}],l}^{(n_2)}(\boldsymbol{g}^\prime)]  {\hat c}_{\boldsymbol{k},n_1}^\dagger {\hat c}_{[\boldsymbol{k}+\boldsymbol{q}],n_2} \nonumber \\
&=& \sum_{\boldsymbol{k}}\sum_{n_1, n_2} F(\boldsymbol{k}_1, [\boldsymbol{k}_1 + \boldsymbol{q}], n_1, n_2, - \boldsymbol{g}_{\boldsymbol{k}_1 + \boldsymbol{q}}) {\hat c}_{\boldsymbol{k}_1, n_1}^\dagger{\hat c}_{[\boldsymbol{k}_1+\boldsymbol{q}], n_2} 
\end{eqnarray}
To make $S({\boldsymbol q})$ dimensionless and normalized, we define $\tilde{S}({\boldsymbol q}) = AS({\boldsymbol q})/N_e$, where $N_e$ is the number of electron. For the main body of the paper and the remainder of this derivation, we refer to $\tilde{S}({\boldsymbol q})$ as $S({\boldsymbol q})$ for convenience. Substituting Eq.~\ref{sup_density2} into Eq.~\ref{sup_Sq} gives:
\begin{eqnarray}\label{sup_Sq2}
S({\boldsymbol q}) &=&  \frac{1}{N_e}  \langle \sum_{\boldsymbol{k}_1} \sum_{n_1, n_2} F(\boldsymbol{k}_1, [\boldsymbol{k}_1 + \boldsymbol{q}], n_1, n_2, - \boldsymbol{g}_{\boldsymbol{k}_1 + \boldsymbol{q}}) {\hat c}_{\boldsymbol{k}_1, n_1}^\dagger{\hat c}_{[\boldsymbol{k}_1+\boldsymbol{q}], n_2} \nonumber \\
&& \times \sum_{\boldsymbol{k}_2} \sum_{n_3, n_4} F(\boldsymbol{k}_2, [\boldsymbol{k}_2 - \boldsymbol{q}], n_3, n_4, -\boldsymbol{g}_{\boldsymbol{k}_2 - \boldsymbol{q}}){\hat c}_{\boldsymbol{k}_2, n_3}^\dagger{\hat c}_{[\boldsymbol{k}_2-\boldsymbol{q}], n_4} \rangle
\end{eqnarray}
By using fermionic anti-commutation relations, the above equation can be separated into two parts:
\begin{eqnarray}\label{sup_Sq3}
S({\boldsymbol q}) &=& \frac{1}{N_e} \sum_{\boldsymbol{k}_1, \boldsymbol{k}_2} \sum_{n_1, n_2, n_3, n_4} F(\boldsymbol{k}_1, [\boldsymbol{k}_1 + \boldsymbol{q}], n_1, n_2, -\boldsymbol{g}_{\boldsymbol{k}_1 + \boldsymbol{q}}) F(\boldsymbol{k}_2, [\boldsymbol{k}_2 - \boldsymbol{q}], n_3, n_4, -\boldsymbol{g}_{\boldsymbol{k}_2 - \boldsymbol{q}}) \nonumber \\
&& \times (\delta_{[\boldsymbol{k}_1 + \boldsymbol{q}], \boldsymbol{k}_2}\delta_{n_2, n_3}\langle {\hat c}_{\boldsymbol{k}_1, n_1}^\dagger{\hat c}_{[\boldsymbol{k}_2-\boldsymbol{q}], n_4} \rangle + \langle {\hat c}_{\boldsymbol{k}_1, n_1}^\dagger{\hat c}_{\boldsymbol{k}_2, n_3}^\dagger {\hat c}_{[\boldsymbol{k}_2-\boldsymbol{q}], n_4}{\hat c}_{[\boldsymbol{k}_1+\boldsymbol{q}], n_2}  \rangle ) \nonumber \\
&=& \frac{1}{N_e} [\langle {\hat S}_1(\boldsymbol q) \rangle + \langle {\hat S}_2(\boldsymbol q) \rangle ]
\end{eqnarray}
Here, ${\hat S}_1(\boldsymbol q)$ and ${\hat S}_2(\boldsymbol q)$ represents the one-body and two-body contribution of $S(\boldsymbol q)$, respectively. $S_2(\boldsymbol q)$ can be written as:
\begin{eqnarray}\label{sup_Sq2}
{\hat S}_2(\boldsymbol q) &=& \sum_{\boldsymbol{k}_1, \boldsymbol{k}_2} \sum_{n_1, n_2, n_3, n_4} F(\boldsymbol{k}_1, [\boldsymbol{k}_1 + \boldsymbol{q}], n_1, n_2, -\boldsymbol{g}_{\boldsymbol{k}_1 + \boldsymbol{q}}) F(\boldsymbol{k}_2, [\boldsymbol{k}_2 - \boldsymbol{q}], n_3, n_4, -\boldsymbol{g}_{\boldsymbol{k}_2 - \boldsymbol{q}}) {\hat c}_{\boldsymbol{k}_1, n_1}^\dagger{\hat c}_{\boldsymbol{k}_2, n_3}^\dagger {\hat c}_{[\boldsymbol{k}_2-\boldsymbol{q}], n_4} {\hat c}_{[\boldsymbol{k}_1+\boldsymbol{q}], n_2} \nonumber \\
\end{eqnarray}
Recall Eq.~\ref{BP_ham} and Eq.~\ref{coulomb_tensor}, we note $S_2(\boldsymbol q)$ and the interaction part of ${\hat H}_{\rm HBP}$ shares the same structure upon a change of variable. Starting from Eq.~\ref{BP_ham}, by using $\boldsymbol{k}_1 = [\boldsymbol{k}_l - \boldsymbol{q}]$ and  $\boldsymbol{k}_2 = [\boldsymbol{k}_k + \boldsymbol{q}]$, we have: 
\begin{eqnarray}\label{sup_Sq2_2}
\langle {\hat H}_{\rm int} \rangle &=& \sum_{\boldsymbol q}\sum_{i,j,k,l} \frac{V({\boldsymbol q})}{2A} F([{\boldsymbol k}_k + {\boldsymbol q}], {\boldsymbol k}_k , n_i, n_k, {\boldsymbol g}_{{\boldsymbol k}_k + {\boldsymbol q}}) F([{\boldsymbol k}_l - {\boldsymbol q}], {\boldsymbol k}_l, n_j, n_l, {\boldsymbol g}_{{\boldsymbol k}_l - {\boldsymbol q}}) \langle c^\dagger_{[{\boldsymbol k}_k + {\boldsymbol q}], n_i}c^\dagger_{[{\boldsymbol k}_l - {\boldsymbol q}], n_j} c_{\boldsymbol{k}_l, n_l}c_{\boldsymbol{k}_k, n_k}\rangle  \nonumber \\
&=&\sum_{\boldsymbol q} \sum_{\boldsymbol{k}_1, \boldsymbol{k}_2} \sum_{n_i,n_j,n_k,n_l} \frac{V({\boldsymbol q})}{2A} F({\boldsymbol k}_2, [{\boldsymbol k}_2 - {\boldsymbol q}] , n_i, n_k, {\boldsymbol g}_{{\boldsymbol k}_k + {\boldsymbol q}}) F({\boldsymbol k}_1, [{\boldsymbol k}_1 + {\boldsymbol q}], n_j, n_l, {\boldsymbol g}_{{\boldsymbol k}_l - {\boldsymbol q}}) \langle c^\dagger_{{\boldsymbol k}_2, n_i}c^\dagger_{{\boldsymbol k}_1, n_j} c_{[{\boldsymbol k}_1 + {\boldsymbol q}], n_l}c_{[{\boldsymbol k}_2 - {\boldsymbol q}], n_k}\rangle  \nonumber \\
&=& -\sum_{\boldsymbol q} \sum_{\boldsymbol{k}_1, \boldsymbol{k}_2} \sum_{n_i,n_j,n_k,n_l}\frac{V({\boldsymbol q})}{2A} F({\boldsymbol k}_1, [{\boldsymbol k}_1 + {\boldsymbol q}], n_j, n_l, -\boldsymbol{g}_{\boldsymbol{k}_1 + \boldsymbol{q}})F({\boldsymbol k}_2, [{\boldsymbol k}_2 - {\boldsymbol q}] , n_i, n_k, -\boldsymbol{g}_{\boldsymbol{k}_2 - \boldsymbol{q}}) \langle c^\dagger_{{\boldsymbol k}_1, n_j} c^\dagger_{{\boldsymbol k}_2, n_i} c_{[{\boldsymbol k}_1 + {\boldsymbol q}], n_l}c_{[{\boldsymbol k}_2 - {\boldsymbol q}], n_k}\rangle  \nonumber \\
&=& -\sum_{\boldsymbol q}\frac{V({\boldsymbol q})\langle {\hat S}_2(\boldsymbol q) \rangle}{2A}
\end{eqnarray}
In the above derivation, we used ${\boldsymbol g}_{{\boldsymbol k}_k + {\boldsymbol q}} = {\boldsymbol k}_k + {\boldsymbol q} - {\boldsymbol k}_2 = - {\boldsymbol g}_{{\boldsymbol k}_2 - {\boldsymbol q}}$, and similarly for ${\boldsymbol g}_{{\boldsymbol k}_l - {\boldsymbol q}}$. Similarly, ${\hat S}_1(\boldsymbol q)$ can be written as:
\begin{eqnarray}\label{sup_S1q}
{\hat S}_1(\boldsymbol q) &=& \sum_{\boldsymbol{k}_1, \boldsymbol{k}_2} \sum_{n_1, n_2, n_3, n_4} \delta_{[\boldsymbol{k}_1 + \boldsymbol{q}], \boldsymbol{k}_2}\delta_{n_2, n_3}F(\boldsymbol{k}_1, [\boldsymbol{k}_1 + \boldsymbol{q}], n_1, n_2, -\boldsymbol{g}_{\boldsymbol{k}_1 + \boldsymbol{q}}) F(\boldsymbol{k}_2, [\boldsymbol{k}_2 - \boldsymbol{q}], n_3, n_4, -\boldsymbol{g}_{\boldsymbol{k}_2 - \boldsymbol{q}}) \langle {\hat c}_{\boldsymbol{k}_1, n_1}^\dagger{\hat c}_{[\boldsymbol{k}_2-\boldsymbol{q}], n_4} \rangle \nonumber \\
&=& \sum_{\boldsymbol{k}_1} \sum_{n_1, n_2, n_4} F(\boldsymbol{k}_1, [\boldsymbol{k}_1 + \boldsymbol{q}], n_1, n_2, -\boldsymbol{g}_{\boldsymbol{k}_1 + \boldsymbol{q}}) F([\boldsymbol{k}_1 + \boldsymbol{q}], \boldsymbol{k}_1, n_2, n_4, \boldsymbol{g}_{\boldsymbol{k}_1 + \boldsymbol{q}}) \langle {\hat c}_{\boldsymbol{k}_1, n_1}^\dagger{\hat c}_{\boldsymbol{k}_1, n_4} \rangle \nonumber \\
&=& \sum_{\boldsymbol{k}_1} \sum_{n_1, n_2, n_4} F(\boldsymbol{k}_1, [\boldsymbol{k}_1 + \boldsymbol{q}], n_1, n_2, -\boldsymbol{g}_{\boldsymbol{k}_1 + \boldsymbol{q}}) F^\ast(\boldsymbol{k}_1,[\boldsymbol{k}_1 + \boldsymbol{q}],  n_4, n_2, -\boldsymbol{g}_{\boldsymbol{k}_1 + \boldsymbol{q}}) \langle {\hat c}_{\boldsymbol{k}_1, n_1}^\dagger{\hat c}_{\boldsymbol{k}_1, n_4} \rangle
\end{eqnarray}
At the limit of $N_b = \infty$, $\langle {\hat S}_1(\boldsymbol q) \rangle = N_e$ due to completeness of Bloch states, which makes $S(\boldsymbol q) = 1 + \langle {\hat S}_2(\boldsymbol q)\rangle/N_e$. Although numerical cutoff would invalidate this relationship, we note that calculation of $S_1(\boldsymbol q)$ is still unnecessary, as it has been proven that $\langle {\hat S}_2(\boldsymbol q)\rangle$ remains unchanged upon band projection\cite{zaklama2024structure}. Therefore, $S(\boldsymbol q)$ can be reconstructed from the truncated two-body contribution $\langle {\bar S}_2(\boldsymbol q)\rangle$ via $S(\boldsymbol q) = 1 + \langle {\bar S}_2(\boldsymbol q)\rangle/N_e$

To evaluate $S(\boldsymbol q)$ under the framework of NQS, we rewrite ${\bar S}_2(\boldsymbol q)$ by inserting resolution of identity:
\begin{eqnarray}\label{sup_S2_sample}
S_{\theta}(\boldsymbol q) = 1+ \frac{\bra{\Psi_\theta}  {\bar S}_2(\boldsymbol q) \ket{\Psi_\theta}}{\bra{\Psi_\theta}\Psi_\theta\rangle} = 1+ \mathbb{E}_{{\boldsymbol x} \sim|\psi_\theta({\boldsymbol x})|^2}[\frac{\sum_{{\boldsymbol x}^\prime}\bra{{\boldsymbol x}} {\bar S}_2(\boldsymbol q)\ket{{\boldsymbol x}^\prime} \psi_\theta({\boldsymbol x}^\prime)}{\psi_\theta({\boldsymbol x})}]
\end{eqnarray}
where ${\bar S}_2(\boldsymbol q)$ is the band truncated version of Eq.~\ref{sup_Sq2}. At $|\boldsymbol q| = 0$, $S_{\theta}(\boldsymbol q) = N_e^2$ by definition. For clearer visualizations, we suppress this divergence by setting $S_{\theta}(0) = 0$.

\section{Details on Neural Network Architectures}
To introduce the architecture of Neural Transformer Backflow(NTB), We start with a bitstring $\ket{\boldsymbol x} = \ket{x_1, x_2, \dots, x_n}$ that represents the basis states of ${\hat H}_{\rm MBP}$, with $x_n \in \{0,1\}$. The bitstring is first embedded into a $N_{\rm{emb}}$-dimensional vector space, augmented with sinusoidal positional encoding, and then passed through a Transformer with $N_{\rm layer}$ layers, $N_{\rm head}$ attention heads and hidden dimension $N_{\rm hid}$\cite{vaswani2017attention}. Altogether, the above transformation can be expressed as:
\begin{eqnarray}\label{transformer layer}
&h_{n, h} = {\rm Transformer}(x)_n&
\end{eqnarray}
where $n = 1,\dots,N_o$ and $h = 1,\dots,N_{\rm hid}$ index the number of orbitals and hidden dimension, respectively. For NTB, the $n^{th}$ orbital's component of the contextual layer is projected onto a set of neural orbitals using linear layers:
\begin{eqnarray}
\phi_{n,m}^{k} = \sum_h [W_{n, m, h, k}^{\rm r}h_{n,h} + i W_{n, m, h, k}^{\rm i}h_{n,h}] + (b_{n,m,k}^{\rm r} + ib_{n,m,k}^{\rm i})
\end{eqnarray}
where $m=1,\dots,N_{\rm e}$ and $k=1,\dots,N_{\rm det}$ index the number of electron and the number determinants, respectively. We choose $N_{\rm layer} = 2$, $N_{\rm head} = 16$, $N_{\rm hid} = 32$, $N_{\rm emb} = 200$ throughout this work. The above architecture can be viewed as a natural extension of the Hartree–Fock (HF) solution. For $h=1$, the $n^{\rm th}$ orbital produces $W_{n,m}$ of shape $N_e$, matching the shape of the mean-field orbitals; contracting $W_{n,m}$ with ${\bf h}_n$ plays the same role as orbital selection in the HF method. When $N_{\rm hid}>1$, the Transformer uses the hidden dimension to encode correlations between qubits. As $W_{n,m,h}$ becomes a learnable many-body orbital, contraction over the hidden dimension acts as a soft selection encoded with many-body information. This significantly enhances the expressiveness of the architecture. 

To encodes more physical information, we select neural orbitals base on the input configuration, which is known as the Backflow selection. By summing over determinants, the output wavefunction amplitude can be written as:
\begin{eqnarray}\label{backflow selection}
\psi_\theta(\boldsymbol x) = \sum_{k}^{N_{\rm det}}\text{det}[\phi_{\{n|x_n=1\},m,k}]
\end{eqnarray}

In the manuscript, we benchmark NTB with two other architectures. As $h_{n, h}$ is generated via Eq.~\ref{transformer layer} for both architectures, the Lattice-Psiformer (L-PFM) architecture based on Ref.~\cite{von2022self} implements:
\begin{eqnarray}
\phi_{i,j,l}^{\rm Pfm} = \sum_h [W_{m, h, k}^{\rm r}h_{n,h} + i W_{ m, h, k}^{\rm i}h_{n,h}] + (b_{n,m,k}^{\rm r} + ib_{n,m,k}^{\rm i})
\end{eqnarray}
where all bits share the same linear layer $W_{m, h, k} \in \mathbb{R}^{N_{\rm hid}\times N_e N_{\rm det}}$. The wavefunction amplitude is then computed via Eq.~\ref{backflow selection}. For the second architecture, auto-regressive evaluation of wavefunction amplitude is directly performed with Transformer output:
\begin{eqnarray}
p_i(m|x_{<i}) = {\rm softmax}\Big(\sum_k[ W_{m,k}^{\rm re}h_{i,k} + iW_{m,k}^{\rm im}h_{i,k}] + (b_m^{\rm re} + i b_m^{\rm im}) \Big) ~~~~\text{with}~~
P(\mathbf x) = \prod_{i=1}^{N_o}p_i(x_i|x_{<i})
\end{eqnarray}
Here, $m = 0, 1$ represents possible values of each bit. The NTB with $N_{\rm det} = 64$ has $6.97 \times 10^{5}$ parameters. To keep the number of parameters around the same level, for L-PFM, we choose $N_{\rm hid} = 32$, $N_{\rm layer} = 2$, $N_{\rm head} = 16$, and $N_{\rm det} = 1000$, which yields $7.21 \times 10^{5}$ parameters. For Transformer, we choose $N_{\rm hid} = 240$, $N_{\rm layer} = 2$, and $N_{\rm head} = 10$, which yields $7.21 \times 10^{5}$ parameters.

\section{AHFL analysis}

\begin{figure}[htbp]
\centering
\centerline{\includegraphics[width=90mm]{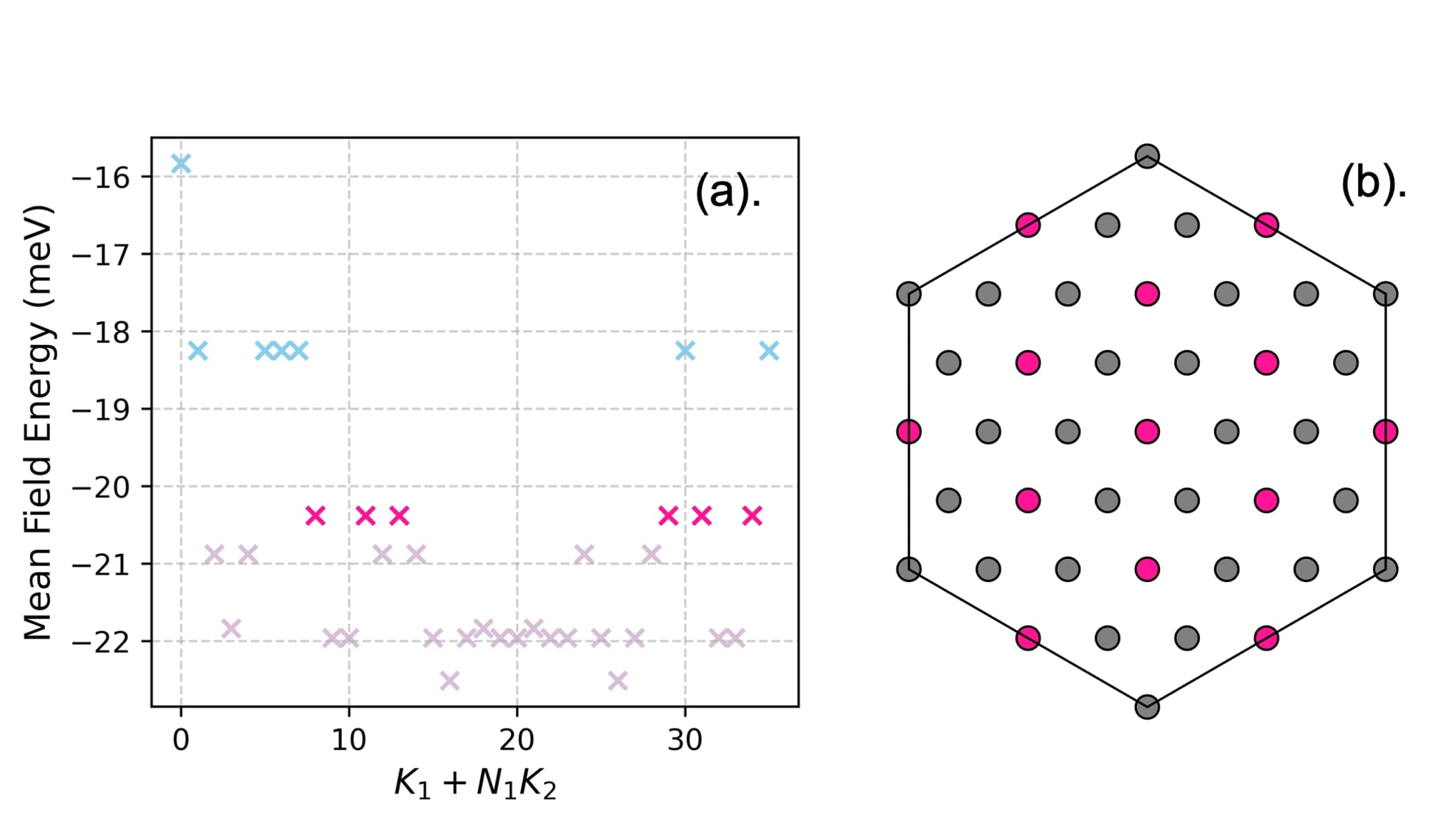}}
\caption{(a). Mean field electron energies obtained by diagonalizing single particle Hamiltonian. Deep pink points represent the $6$ degenerate highest occupied orbitals. (b). Ground state momentum sectors (colored pink) predicted by the fermi liquid theory.}
\label{AHFL}
\end{figure}

For the Anomalous Hall Fermi Liquid (AHFL) state at $\nu = 3/4$, in Fig.~\ref{AHFL}a, the single particle energies of the system at each momentum sector are plotted. A $6$-fold degeneracy is found at $K_1 + N_1K_2 = 8,~11,~13,~29,~31,~34$ between $E_{23}$ and $E_{29}$. With $N_e = 27$, according to the fermi liquid theory, $4$ electrons will occupy the $6$ degenerate orbitals, giving $C_4^6 = 15$ degenerate many-body states. It is found that these states occupies $10$ unique CM momentum sectors at $K_1 + N_1K_2 = 0(\Gamma),~3(M_1),~8, ~11, ~13, ~18(M_2), ~21(M_3), ~29, ~31, ~34$, which has been shown in Fig.~\ref{AHFL}b. This confirms with the NTB calculation, where degeneracies in ground state energies are found in $\Gamma$ and $M$ points.

\section{Details on Optimization and Sampling}
To improve the optimization of neural network, we use the following strategies to avoid local minima during training.

\begin{itemize}
\item{Sign Gradient.} We adopt the signSGD method to update network parameters\cite{bernstein2018signsgd}:
\begin{eqnarray}
\theta_{k+1} = \theta_{k} - {\rm lr} \times {\rm sign}[\nabla_{\theta} E_\theta]
\end{eqnarray}
where $k$ is the optimization step.

\item{Multi-step learn rate.} We employ a multi-step learning rate scheduler with milestones at steps 100, 500, 1000, 1800, 2500, 4000, 6000 and 8000. The initial learning rate is $5 \times 10^{-3}$ with a decay factor $\gamma = 0.5$.
\item{Transfer learning. }We employ transfer learning to initialize optimization in related systems. Because systems from different symmetry sectors or with varying model parameters share the same parameter structure, we find transfer learning between them to be highly effective.
\item{Random perturbation. } We add randomly generated noise to the parameters during the early stages of training to help escape local minima. The noise scale is selected such that the perturbation raises the energy by approximately $0.5$ meV.
\end{itemize}

To impose momentum constrain on MCMC sampling, for a sample $\ket{\boldsymbol x} = \ket{x_1, x_2, \dots, x_n}$, with $x_n \in \{0,1\}$, we create a momentum conserved flip set:
\begin{eqnarray}
S &=& \{ (i,j;k,l) | x_i,x_j = 0; x_k, x_l = 1; [\boldsymbol{k}_i + \boldsymbol{k}_j] = [\boldsymbol{k}_k + \boldsymbol{k}_l] \}
\end{eqnarray}
Proposals are then drawn randomly from $S$ and accepted based on the Metropolis–Hastings algorithm. We perform $N_o$ MCMC flips between successive parameter updates to ensure decorrelation of the Markov chain.

\end{document}